\documentclass[twocolumn]{revtex4}
\usepackage{eurosym}
\usepackage{graphicx}
\usepackage{amssymb}

\def\beq{\begin{equation}}
\def\eeq{\end{equation}}
\def\bea{\begin{eqnarray}}
\def\eea{\end{eqnarray}}
\def\ba{\begin{array}}
\def\ea{\end{array}}
\def\bit{\begin{itemize}}
\def\eit{\end{itemize}}

\def\om{\omega}

\makeatletter
\def\underbracket{  \@ifnextchar [     {\@underbracket}    {\@underbracket [\@bracketheight]}}
\def\@underbracket[#1]{  \@ifnextchar [     {\@under@bracket[#1]}    {\@under@bracket[#1][0.4em]}}
\def\@under@bracket[#1][#2]#3{  \mathop {    \vtop {      \m@th \ialign 
{        ##\crcr $\hfil \displaystyle {#3}\hfil $       \crcr \noalign        {\kern 3\p@ \nointerlineskip }        \upbracketfill {#1}{#2}
       \crcr \noalign        {\kern 3\p@ }     }   }  }  \limits}
\def\upbracketfill#1#2{  $\m@th \setbox \z@ \hbox {$\braceld$}
  \edef\@bracketheight{\the\ht\z@}\bracketend{#1}{#2}
  \leaders \vrule \@height #1 \@depth \z@ \hfill
  \leaders \vrule \@height #1 \@depth \z@ \hfill  \bracketend{#1}{#2}$}
\def\bracketend#1#2{\vrule height #2 width #1\relax}

\begin{document}

\title{Electron pairs' sliding states in superconductivity}
\author{Krzysztof A. Meissner$^1$ and Lucjan Piela$^2$}
\affiliation{$^1$Faculty of Physics, University of Warsaw\\
ul. Pasteura 5, 02-093 Warsaw, Poland\\
$^2$Faculty of Chemistry, University of Warsaw\\
ul. Pasteura 1, 02-093 Warsaw, Poland}
\email{piela@chem.uw.edu.pl}

\begin{abstract}
A quantum-mechanical model of Cu-O-Cu-O four-center four-electron part of
the copper-oxide plane embedded in a superconducting crystal La$_{2-x}$Sr$%
_{x}$CuO$_{4}$ (LSCO) is considered. It is shown that displacing the
nearest-neighbor La(Sr)O plane lattice atoms off by a distance as small as $%
\sim $ $\pm 0.1$~\AA, i.e. of the order of ground state
vibrations dictated by the Heisenberg uncertainty principle, may trigger a
dramatic change of the ground state electronic charge distribution in the
CuOCuO system. This results in the electron pairs' concerted sliding within
the system over distances about 2 \AA , i.e. close to the copper oxygen
atomic distance. The effect depends crucially on the lattice crystal field
and doping. The appearance of energy gaps associated with the electronic
states' avoided crossings points to a universal nature of the phenomenon.
The results suggest a generalization of the models used up to now in 
description of strongly correlated electrons.
\end{abstract}

\maketitle

%\pacs{....................}

After 34 years from the discovery by Bednorz and Miller the mechanism behind
the high-temperature superconductors (HTS) is still unclear (for a recent
pedagogical review see \cite{revHTS}). There is a consensus that strong
electronic correlation is to be taken into account \cite{Dagotto} but the
goal of finding and solving the proper model is not yet achieved (even
though the magnetic properties are probably sufficiently well described by
the Hubbard \cite{Hubbard} or $t-J$ \cite{Spalek} models \cite{KW}). On the
other hand the highest critical temperatures for cuprates correspond to a
definite CuO distance equal to $1.923$ \AA\ - the discovery virtually
unnoticed by the HTS community (\cite{FG}, confirming an earlier less
accurate observation \cite{RG}), indicates a sharp electronic instability
with respect to lattice distances. The instability has been suggested first
by Burdett \cite{Burdett} as an avoided crossing point of two hypothetical
electronic states (a result of a low-energy conical intersection, see e.g. 
\cite{Piela}). The present letter shows by explicit calculation the
importance of ground state lattice vibrations on the presence of
avoided crossings of electronic states.

Stable atomic and molecular structures represent a subject of most quantum
chemical investigations. In the present letter the target is quite the
opposite: to apply theory to find such an unstable electronic ground-state
that, under some small changes of the nuclear framework, may produce
dramatically different electronic density distribution with a kind of
electronic pairs sliding (`sliding states') - a prerequisite of the
resistance-free electron pairs' flow.

Although the model presented is general we chose to focus on a particular
lattice as an example, here taken as one of the first HTSs discovered: the La%
$_{2-x}$Sr$_{x}$CuO$_{4}$ crystal (LSCO), where $x$ denotes the doping
level. The model pertains to a fragment of the LSCO crystal, Fig.1. 
\begin{figure}[t]
\centering
\includegraphics[width=7.8truecm,viewport= 0 0 800 600,clip]{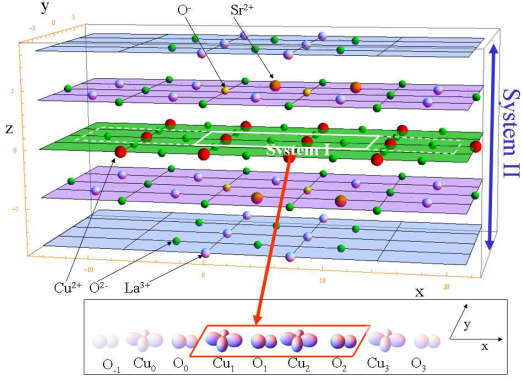}
\caption{The LSCO crystal fragment considered: system I treated
quantum-mechanically is embedded in the electrostatic field of system II. In
system I, the inner electronic shells of an atom together with its nucleus,
are modeled as a spherical Gaussian pseudo-nucleus, with the exponent
coefficients $0.577$ u and $0.115$ u for the oxygen and copper atoms,
respectively (the corresponding mean radii are chosen to be equal to the
atomic ones, the core charges $1.05$\ and $0.95$\ are used for the copper
and the oxygen atoms, respectively. The $3d_{x^{2}-y^{2}}^{\mathrm{Cu}}$\
and $2p_{x}^{\mathrm{O}}$\ orbitals are taken as composed of Gaussian type
spherical lobes, with the nucleus-lobe distance equal to $0.5$\ u and the
lobe's exponent values: $a_{3d}=0.04$ and $a_{2p}=0.16$ for which the effect
is most pronounced. The model's calculated electronic states
depend in an important way on the lattice ions' displacements in system II
(off their lattice positions) $d\equiv d_{\mathrm{O}}=d_{\mathrm{La}}$.  In
order to be on the safe side we have reduced this coupling by a factor of ${%
\frac{1}{2}}$ and modeled this effect by $\Delta a_{3d}/a_{3d}=$ $0.15\tanh
\left( 10d_{\mathrm{O}}\bar{q}_{\mathrm{Cu}}\right) $ and $\Delta
a_{2p}/a_{2p}=0.30\tanh \left( 10d_{\mathrm{La}}\bar{q}_{\mathrm{O}}\right) ,
$ where $\bar{q}_{\mathrm{Cu}}$ and $\bar{q}_{\mathrm{O}}$ are the mean
values of the nearest-neighbor ions' charge for the copper and the oxygen
atoms of system I, respectively.\ The nominal ionic charges of system II are
scaled by $\protect\eta =\frac{1}{3}$. }
\end{figure}

An \textit{ab initio} Valence Bond (VB) method \cite{Shaik} of solving the
Schr\"{o}dinger equation for system I embedded in the LSCO crystal field
produced by system II is applied, Fig.1 (the model allows also to mimic the
doping by adding, removing, replacing and/or shifting some surrounding
lattice ions of system II). The wave functions and energies are obtained by
using Ritz variational approximation, with $N=10$ normalized and mutually
orthogonal resonance (or diabatic) structures $\phi _{i}$ forming the basis
set, $\psi _{n}=\sum_{i=1}^{N}c_{ni}\phi _{i}$:

\begin{eqnarray}
\phi _{1} &=&\left\vert {Cu}{\uparrow \downarrow }\ {O}\ {Cu}{\uparrow
\downarrow }\ {O}\right\rangle \ \mathrm{ionic\ CuCu}  \nonumber \\
\phi _{2} &=&\left\vert {Cu}\ {O}{\uparrow \downarrow }\ {Cu}\ {O}{\uparrow
\downarrow }\right\rangle \ \mathrm{ionic\ OO}  \nonumber \\
\phi _{3} &=&\left\vert {Cu}\ {O}{\uparrow \downarrow }\ {Cu}{\uparrow
\downarrow }\ {O}\right\rangle \ \mathrm{ionic\ OCu}  \nonumber \\
\phi _{4} &=&\left\vert {Cu}{\uparrow \downarrow }\ {O}\ {Cu}\ {O}{\uparrow
\downarrow }\right\rangle \ \mathrm{ionic\ Cu...O}  \nonumber \\
\phi _{5} &=&\left\vert {Cu}{\downarrow }\ {O}{\uparrow }\ {Cu}{\downarrow }%
\ {O}{\uparrow }\right\rangle \ \mathrm{antiferromagnetic\ 1}  \nonumber \\
\phi _{6} &=&\left\vert {Cu}{\uparrow }\ {O}{\downarrow }\ {Cu}{\uparrow }\ {%
O}{\downarrow }\right\rangle \ \mathrm{antiferromagnetic\ 2}  \nonumber \\
\phi _{7} &=&\left\vert {Cu}{\uparrow \downarrow }\ {O}{\downarrow }\ {Cu}{%
\uparrow }\ {O}\right\rangle \ \mathrm{ionic\ Cu/antiferro1}  \nonumber \\
\phi _{8} &=&\left\vert {Cu}{\uparrow \downarrow }\ {O}{\uparrow }\text{ }{Cu%
}{\downarrow }\ {\small O}\right\rangle \ \mathrm{ionic\ Cu/antiferro2} 
\nonumber \\
\phi _{9} &=&\left\vert {Cu}\ {O}{\downarrow }\ {Cu}{\uparrow }\ {O}{%
\uparrow \downarrow }\right\rangle \ \mathrm{antiferro1/ionic\ O}  \nonumber
\\
\phi _{10} &=&\left\vert {Cu}\ {O}{\uparrow }\ {Cu}{\downarrow }\ {O}{%
\uparrow \downarrow }\right\rangle \ \mathrm{antiferro2/ionic\ O.}
\end{eqnarray}

The $i^{\mathrm{th}}$\ resonance structure $\phi _{i}$ represents a
normalized Slater determinant built of four orthonormal atomic spinorbitals.
To minimize the termini effects the symmetric orthogonalization \cite%
{Loewdin} has been performed for nine atomic orbitals (AOs): $%
3d_{x^{2}-y^{2}}^{\mathrm{\ Cu}}$ (for copper atoms, $xy$ is a $\mathrm{CuO}%
_{2}$ plane) and $2p_{x}^{\mathrm{O}}$ (for oxygen atoms) centered,
respectively, along the $\mathrm{O}_{-1},\mathrm{Cu}_{0},\mathrm{O}_{0},%
\mathrm{Cu}_{1},\mathrm{O}_{1},\mathrm{Cu}_{2},\mathrm{O}_{2},\mathrm{Cu}%
_{3},\mathrm{O}_{3}$ axis ($x$), see the inset of Fig.1. Due to the
symmetric orthogonalization the resulting orthogonal orbitals (OAOs) have
the least possible deviation from the starting AOs. The four OAOs that are
to be used to build $\phi _{i}$ functions\ are centered on four consecutive
Gaussian spherical atomic cores symbolized by the centers $\mathrm{Cu}_{1},%
\mathrm{O}_{1},\mathrm{Cu}_{2},\mathrm{O}_{2}$ (system I).

The present approach uses a superposition of the diabatic structures to
describe the electron transfer, and thus, in a general sense, goes along
with the heuristic Marcus two-parabolas picture of electron transfer \cite%
{Marcus} and its three-parabolas extensions by Larsson \cite{Larsson}.  
The electronic Hamiltonian ($\hat{H}$) of system I contains all the
(non-relativistic) terms: the kinetic energy and Coulombic interaction
operators of the four electrons with themselves, with the cores and the core-core-interaction
of system I, as well as four electrons with 102 point-like ions of system II. The model presented
(when modified with other types of AOs) shows the sliding effect to appear
also for the $\pi $ states coming from the $3d_{xy}^{\mathrm{Cu}}$ and $%
2p_{y}^{\mathrm{O}}$ interaction, as well as for the interacting $1s$ - type
orbitals ($\sigma $ state). 
\begin{figure}[t]
\centering
\includegraphics[width=8.6truecm,viewport= 0 0 720 580,clip]{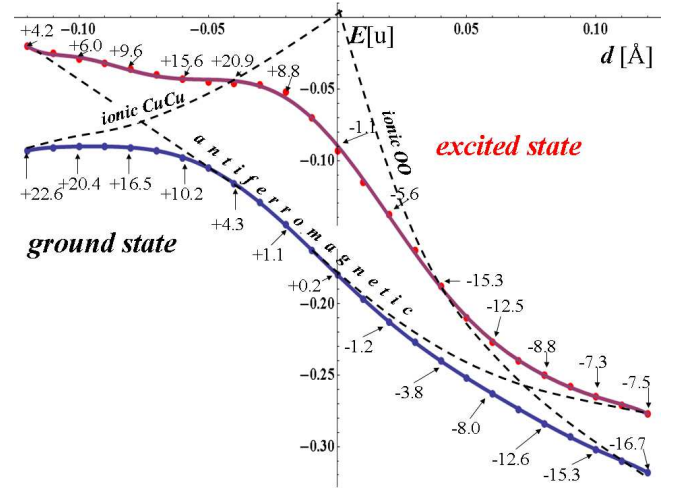}
\caption{The energies (parameters as in the text) of the
ground- and the first-excited electronic singlet states as functions of the
ions' lattice displacement $d$. Along the curves the
corresponding values of the dipole moment are given (in Debye units, D). The
values of the dipole moments reflect the presence of the avoided crossings:
between the highly polarized ionic CuCu and low-polarity antiferromagnetic
states (left) and between the antiferromagnetic and the oppositely polarized
ionic OO states (right).}
\end{figure}

\begin{figure}[t]
\centering
\includegraphics[width=7.8truecm,viewport= 0 0 700 540,clip]{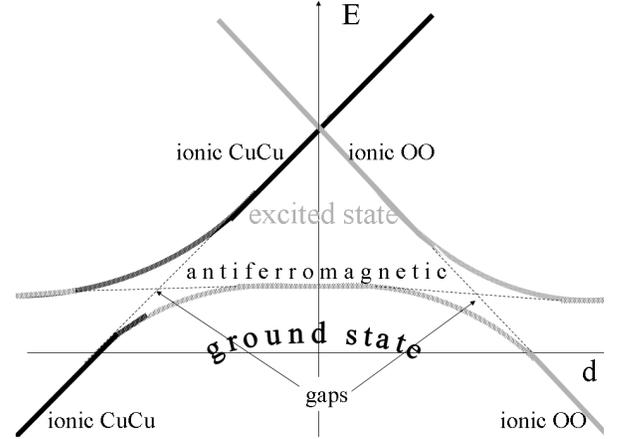}
\caption{Schematic representation (`letter A') of the ground and excited
states illustrating electron pairs' sliding states}
\end{figure}

Fig. 2 and schematic Fig. 3 show the dependence of the ground- and the first excited
states' electronic energies as functions of $d=d_{\mathrm{La}}(=d_{\mathrm{O}%
})$. As we can see the ions' displacements of system II may lower the energy
of an excited large-polarity state to such an extent that it becomes the
ground state, instead of the low-polarity and usually lowest-energy
antiferromagnetic state, thus changing profoundly the character of the
ground state. It should be emphasized, as explained below, that the
displacements needed (about 0.1 \AA ) are of the order of the \textit{ground
state} amplitude of vibrations of the lattice. The two avoided crossings
gaps appear, $\Delta _{\mathrm{cross1}}$ and $\Delta _{\mathrm{cross2}}$;
for the parameters assumed in this letter they are equal to $\Delta _{%
\mathrm{cross1}}\approx $ $1.5$ eV and $\Delta _{\mathrm{cross2}}\approx 0.92
$ eV.

The calculated ground-state electron density distribution $\rho _{0}\left( {%
\mathbf{r}}\right) =$ $\left\langle \psi _{0}|\Sigma _{i=1}^{4}\delta ({%
\mathbf{r}}-{\mathbf{r}}_{i})\psi _{0}\right\rangle $, ${\mathbf{r}}_{i}$
being the $i^{\mathrm{th}}$ electron position, as a function of the ions'
displacement $d,$ shows the details of the two quite abrupt changes at two
consecutive avoided crossings during a vibration of a certain amplitude
corresponding to a long-distance concerted sliding\ of the two interacting
electron pairs by about the CuO distance $1.92$ \AA .

The model used allows to study doping effects. At a chosen doping $x\sim
0.125$ (in an example, Fig.1: two Sr ions replacing two La ions, with
appropriate changes in charges of two O ions to keep charge neutrality) and
optimizing orbital parameters the two avoided crossings have been within a
reasonable vibrational amplitude. If the doping have been chosen bigger then
the avoided crossings are still present but the $d_{\mathrm{cross}}$ values
shift by about a hundredth of \AA\ per an added/removed dopant atom.
Therefore, if initially the values of $d_{\mathrm{cross}}$ correspond to the
physical range of the ions' vibrations, the doping exceeding some degree may
lead to going outside of an effective vibrationally-allowed $d$. The
disappearance of the sliding effect with the excessive doping explains two
puzzles: why doping is necessary for cuprates to obtain a HTS and why there
exists an optimal doping. On the other hand the fact that
the energy levels in the CuO$_{2}$ plane depend very little on the oxygen mass
and only parametrically on the
vibrations of the La(Sr)O plane may explain the observation that the isotopic
effect in HTS is very small, in distinction to the BCS theory.

It is also important to note that the vibrations of the lattice, crucial for
the proposed effect, are for $T\rightarrow 0$ given by (an average
displacement squared in one direction of a single atom of mass $M$ {in a
crystal in the Debye theory with Debye temperature }$\theta _{D}$ \cite%
{Ziman}) 
\begin{equation}
\langle \Delta z^{2}\rangle =\frac{3\hslash ^{2}}{4Mk_{B}\theta _{D}}\left(
1+\frac{2\pi ^{2}T^{2}}{3\theta _{D}^{2}}+O\left( \frac{T}{\theta _{D}}%
\mathrm{e}^{-\theta _{D}/T}\right) \right) .
\label{gsvibrations}
\end{equation}%
As we see, the amplitudes of vibrations are dominated by ground
state vibrations (existing for $T=0$ as a result of the Heisenberg
uncertainty principle) and the thermal excitations start to be important
only at high temperatures, comparable to $\theta_D$. These ground state
vibrations have larger amplitudes for smaller Debye temperature. In view of
the effect described in this letter, it appears not accidental that the best
known HTS materials have lattices based on elements with very low Debye
temperatures -- mercury ($72$ K), thallium ($78$ K), barium ($111$ K),
bismuth ($120$ K) or lanthanum ($135$ K). It is consistent with the fact
that experimentally, and unlike the BCS theory, the critical temperature for
the HTSs increases with decreasing Debye temperature \cite{HL}. Plugging in
the numbers for lanthanum oxide we get $\sqrt{\langle \Delta z^{2}\rangle
_{0}}\sim 0.06\ \mathring{A}$ for the ground state average displacements and
it is approximately the value where the electronic ground state of system I
changes its character from antiferromagnetic to ionic as seen in Fig.2.

Since the dipole moments can be very large (up to $\sim \pm $ $20$ D) the
dipole-dipole interactions may make the electron sliding mechanism to become
a coherent motion of many electron pairs within the CuO$_{2}$ planes. The
motion might spread throughout the whole lattice, especially along the
dipole alignment lines ('stripes'). At a given temperature, it may be viewed
as a frustrating competition of the charge density waves (CDW) vs the spin
density waves (SDW). Such a spontaneous polarization \cite{Stolarczyk} would
additionally lower the ground-state energy, thus increasing the gap by $%
\Delta _{\mathrm{coh}}$.

The method used in this letter has several differences with respect to usual
approaches. The first difference comes from the VB method's real-space
material-dependent insight into the electronic correlation: the electron
pairs' formation, interaction and dissociation, also with appearance of
chemical bonds. Secondly, in the tight-binding and band structure pictures
crystal orbitals describe only the exchange part of the electron-electron
correlation and take into account only the ground state. The question which
model should be used within the latter approach to simulate the mechanism of
superconductivity described in this letter requires a comment. In the
undoped case (antiferromagnetic) the presence of ionic states is irrelevant
since they correspond to too large amplitudes of the LaO plane oscillations
and then we expect that the Heisenberg-like model should adequately describe
the case. In the appropriately doped case both the Cu and O orbitals should
be taken into account and then we should use the Emery three-band model (%
\cite{Emery,ES}). However, the effect described in this letter shows that
the parameters used in the model (or its simplified versions) should depend
on the position of La(Sr)O planes and even the most general models
considered up to now did not take into account the time dependence of
parameters induced by the La(Sr)O oscillations. It follows that the
effective models aiming to describe HTS should be generalized to time
dependent (oscillating) parameters. In the three-band model the mechanism
described in this letter could be approximately simulated by a
generalization in which the Coulomb repulsion terms on oxygen $U_{p}$ and on
copper $U_{d}$ (treated in the previous models as constants) depend on time with the
opposite phase, schematically 
\begin{equation}
U_{d}(t)\sim U_{p}^{0}+\Delta _{p}\cos (\omega _{s}t),\ \ \ U_{p}(t)\sim
U_{d}^{0}-\Delta _{d}\cos (\omega _{s}t).
\end{equation}%
$\omega _{s}$ is related to oscillation frequencies in the La(Sr)O plane and
can be estimated from the Heisenberg uncertainty principle  $\sqrt{\Delta z^{2} \Delta p_z^{2}}\ge
\hslash/2$ as 
\begin{equation}
\omega _{s}\sim \frac{\hslash }{2M\langle \Delta z^{2}\rangle }\sim 1.3\cdot
10^{13}\ \mathrm{s}^{-1}.
\end{equation}%
for the lanthanum atom mass and $\sqrt{\Delta z^{2}}\approx 0.06$ \AA 
(from (\ref{gsvibrations}) we see that $\om_s= 2k_B \theta_D/(3\hslash)$). $%
U_{d}^{0}$ and $U_{p}^{0}$ are estimated as 5-6 eV and 2-3 eV, respectively 
\cite{Emery}. The depths of the oscillation $\Delta $'s depend on the doping
-- they can be estimated as half of the gaps  $\Delta
_{\mathrm{cross1}}$ and $\Delta _{\mathrm{cross2}}$ 
so for our case of $x=0.125$ it gives $\Delta _{d}\approx
0.8$ eV and $\Delta _{p}\approx 0.5$ eV. One could also add the dependence
of the hopping parameters $t$ on time with frequency $\omega _{c}$ related
to the oscillation frequencies in the CuO plane. Such time dependences
required by the presented mechanism may significantly change the properties
of the models. As mentioned earlier, because of very large electric dipole
moments in ionic states, one should also add dipole-dipole interactions
extending over larger distances than just nearest neighbors.

The described mechanism indicates a possible general nature of the electron
pairs' sliding appearance whenever the electronic instability (avoided
crossing) can be reached by a suitable choice of materials (elements,
lattice structure, doping) and external conditions (like pressure). It leads
to a necessity, in the three-band model or its simplified versions, to
introduce time dependence of the parameters. The mechanism is also an
example of the quantum ground state vibrations (of the La(Sr)O plane)
coupled to a quantum system (electronic states in the CuO planes) strongly enough
to significantly modify its properties. In conclusion, electron pairs'
sliding states described in this letter may possibly provide  an explanation of
the presence of high temperature superconductivity in copper-oxide planes surrounded by very
special lattice structures.\\

\noindent\textbf{Acknowledgements}

We are very indebted to Wojciech Grochala, Jerzy {\L }usakowski, J\'{o}zef
Spa\l ek, Leszek Stolarczyk and Krzysztof Wohlfeld for stimulating
discussions. K.A.M was partially supported by the Polish National Science
Center HARMONIA project UMO-2015/18/M/ST2/00518.

\end{document}